\documentstyle[aps,epsfig]{revtex}                  
\begin{document}
\pagestyle{empty}                                      
\preprint{
\font\fortssbx=cmssbx10 scaled \magstep2
\hbox to \hsize{
\hfill$\raise .5cm\vtop{              
                \hbox{NCTU-HEP-0003}}$}}
\draft
\vfill
\title{A new look at the pair-production width 
in a strong magnetic field }
\author{W. F. Kao, Guey-Lin
Lin and Jie-Jun Tseng} %
\address{Institute of Physics, National Chiao-Tung University,
Hsinchu 300, Taiwan}
\date{\today}
%
%
\vfill
\maketitle
\begin{abstract}
We reexamine the process $\gamma\to e^++ e^-$ in a background magnetic field comparable to $B_c\equiv m_e^2/e$. 
This process is known to be non-perturbative in the magnetic-field strength. However,
it can be shown that the {\it moments} of the above pair production
width are proportional to the derivatives of photon polarization 
function at the zero energy, which is perturbative in $B$. Hence, the 
pair-production width can be easily obtained 
from the latter by the inverse 
Mellin transform. The implications of our approach are discussed.     
\end{abstract}
%
%
%
%
\pagestyle{plain}
The electroweak phenomena associated with an intensive background magnetic 
field are rather rich. Under a background magnetic field, 
a physical photon can 
decay into an $e^+e^-$ pair or split into two photons. Such processes are
relevant to the attenuation of gamma-rays from pulsars\cite{STU,BH}. 
The study of 
pair production process $\gamma+B\to e^+e^- +B$ was initiated by Toll
\cite{toll} 
long time ago. He obtained a rather tedious expression for the absorption
coefficient $\kappa_{\parallel,\bot}$, where $\parallel$ and $\bot$ denote 
the photon-polarization directions which are respectively 
parallel and perpendicular 
to the plane spanned by the magnetic field ${\bf B}$ and 
the photon momentum ${\bf q}$.
Writing 
\begin{equation}
\kappa_{\parallel,\bot}={\alpha\over 2}\sin\theta\left({eB\over m_e}\right)
T_{\parallel,\bot}(\lambda),
\end{equation}
with $\lambda={3\over 2}(eB/m_e^2)(\omega/m_e)\sin\theta$, Toll obtained
\begin{equation}
T_{\parallel,\bot}(\lambda)={9\over \lambda}
\int_{(6/\lambda)^{2/3}}^{\infty}dv\left[
{-(1-3\eta_{\parallel,\bot}/2\lambda v^{3/2})A'(v)\over 
v^{5/4}(v^{3/2}-6/\lambda)^{1/2}}+
{(v^{3/2}-6/\lambda)^{1/2}\over 3v^{3/4}}A(v)\right], 
\end{equation}
where $\eta_{\parallel}=1$, $\eta_{\bot}=3$ and
\begin{equation}
A(v)={1\over 2\pi}\int_{-\infty+i\epsilon}^{\infty+i\epsilon}
dt e^{ivt+it^3/3},
\end{equation}
where $\omega$ is the photon energy and $\theta$ is the angle 
between the magnetic-field direction and the direction of photon propagation.
We note that, due to the quantization of electron and positron orbits 
in the magnetic field, $\kappa_{\parallel,\bot}$ should contain sawtooth
absorption edges. However, for $B\ll B_c\equiv m_e^2/e$ and 
$\omega \sin\theta \gg 2m_e$, these absorption
edges are rather closely spaced. Hence it is sensible to define an
averaged absorption coefficient which is precisely the $T_{\parallel,\bot}$
(with a trivial prefactor) displayed above. In other words, Toll's result 
is valid for $B \ll B_c$ and $\omega\sin\theta \gg 2m_e$.   
It is interesting to note that, as pointed out by Toll, 
the functions $T_{\parallel,\bot}$ can not be 
calculated order by order in $eB$. 
It is essential to use the exact 
Dirac wave functions for electrons and positrons in the magnetic field
such that the resulting $T_{\parallel,\bot}$ are nonvanishing. The 
non-analytic behaviors of $T_{\parallel,\bot}$ at $eB=0$ can be easily
seen from its asymptotic expression for $\lambda\ll 1$:
\begin{equation}
T_{\parallel,\bot}\to \sqrt{{3\over 2}}({1\over 2}, 
{1\over 4})e^{-4/\lambda}
\end{equation}  
The expression for $T_{\parallel,\bot}$ was simplified considerably 
in the work by Tsai and Erber\cite{TE}. The authors computed the photon
polarization function by the proper-time technique\cite{SCH} and determined
the absorption coefficient $\kappa_{\parallel,\bot}$ using the 
optical theorem. They obtained
\begin{equation}
T_{\parallel,\bot}(\lambda)={4\sqrt{3}\over \pi \lambda}
\int_0^1 dv (1-v^2)^{-1}\left[(1-{1\over 3}v^2), ({1\over 2}+{1\over 6}v^2)
\right]K_{2/3}\left({4\over \lambda}{1\over 1-v^2}\right),
\label{bessel}
\end{equation}     
where $K_{2/3}$ is the modified Bessel function. At the first glance, the
result of Tsai and Erber appears very different from Toll's result. However,
by computing the {\it moments} of $T_{\parallel,\bot}$, the former authors
were able to show that their result is in fact equivalent to that of 
Toll\footnote{To state it more precisely, Tsai and Erber computed the 
{\it moments} of the averaged function $T\equiv 1/2\cdot (T_{\parallel}+
T_{\bot})$, which is relevant to the attenuation of unpolarized photons.}.
We observe that Tsai and Erber simply utilized the {\it moments} of
$T_{\parallel,\bot}$ as a mathematical tool to show the equivalence between
two sets of absorption coefficients. The physical significance of 
these {\it moments} was not studied. In this note, we shall clarify
the meaning of these moments and develop a new method 
of computing the absorption coefficients. Since our approach is essentially 
a systematic expansion in $B/B_c$, it will remain valid for a background 
magnetic field comparable to $B_c$. 

We are motivated by the following contour integral, 
which resembles to the contour integral encountered
in the QCD sum rule calculation of $e^+ e^-\to $ hadrons\cite{SVZ}:
\begin{equation}
I_n=\int_C {d\omega^2\over 2\pi i}{\Pi_{\parallel,\bot}(\omega^2)
\over (\omega^2+\omega_0^2)^{n+1}},
\end{equation}     
where the contour of integration $C$ is shown in Fig. 1.

\vspace{5cm}
\begin{figure}
  \unitlength 1mm
   \begin{center}
      \begin{picture}(25,100)
     \put(-70,30) {\epsfig{file=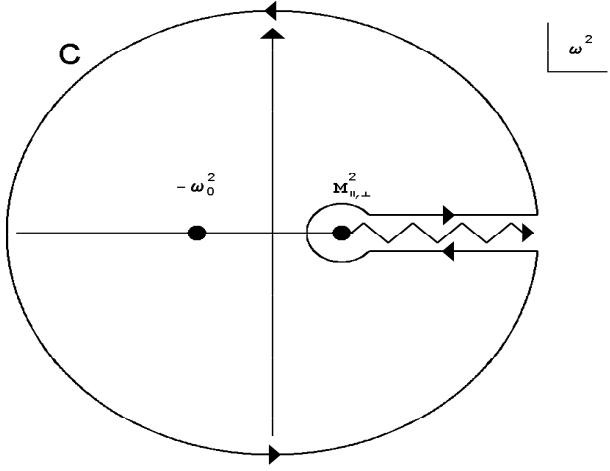,width=10cm,height=10cm}}
      \end{picture}
   \end{center}
\vspace{-3cm}
\caption{The integration contour for $I_n$ and the analytic structure
of $\Pi_{\parallel,\bot}$. In actual calculations, we 
take the radius of the circle to infinity.}
\label{fig1}
\end{figure}
The functions $\Pi_{\parallel,\bot}$ are defined as
\begin{equation}
\Pi_{\parallel,\bot}=\epsilon^{\mu}_{\parallel,\bot}\Pi_{\mu\nu}
\epsilon^{\nu}_{\parallel,\bot},
\end{equation}
where $\epsilon^{\mu}_{\parallel}$ and  $\epsilon^{\mu}_{\bot}$ are 
respectively 
the photon polarization vectors parallel and perpendicular
to the plane spanned by the photon momentum ${\bf q}$ and the magnetic field 
${\bf B}$. We note that the integral $I_n$ may be evaluated in two different ways. One 
computes $I_n$ either by the residue theorem or by a direct integration 
along 
the contour $C$ with the realization that the contribution 
from the outer circle vanishes.  
The equivalence of two integration procedures gives rise to the
relation:
\begin{equation}
{1\over n!}\left({d^n\over d(\omega^2)^n}
\Pi_{\parallel,\bot}\right)\Big{\vert}_{\omega^2=-\omega_0^2}
={1\over \pi}\int_{M^2_{\parallel,\bot}}^{\infty}d\omega^2
{{\rm Im}\Pi_{\parallel,\bot}(\omega^2)\over (\omega^2+\omega_0^2)^{n+1}},
\label{dispersion}
\end{equation} 
where $M_{\parallel,\bot}$ are the threshold energies of pair productions\cite{toll,adl} given by
\begin{equation}
M^2_{\parallel}\sin^2\theta=4m_e^2, \;  
M^2_{\bot}\sin^2\theta=m_e^2\left(1+\sqrt{1+2{B\over B_c}}
\right)^2,
\end{equation} 
with $\theta$ the angle between the photon momentum and the magnetic field.
Since $\kappa_{\parallel,\bot}={\rm Im}\Pi_{\parallel,\bot}/\omega$ 
by the optical theorem, the above equation relate the real part of vacuum
polarization function to the absorption coefficient.

We observe that the l.h.s. of Eq. (\ref{dispersion}) can be easily 
calculated at $\omega^2=-\omega_0^2=0$, since the threshold behaviors
of $\Pi_{\parallel,\bot}$ are absent at this energy value.  
With this choice of $\omega_0^2$,
we recast Eq. (\ref{dispersion}) into
\begin{equation}
{1\over n!}\left({d^n\over d(\omega^2)^n}
\Pi_{\parallel,\bot}\right)\Big{\vert}_{\omega^2=0}
={M_{\parallel,\bot}^{1-2n}\over \pi}\int_{0}^{1}dy\cdot y^{n-1}
\cdot \left(\kappa_{\parallel
,\bot}(y) y^{-1/2}\right).
\label{mellin}
\end{equation}
with $y=M^2_{\parallel,\bot}/\omega^2$.
One notes that the imaginary part of $\Pi_{\parallel,\bot}(\omega^2)$ vanishes for the range $0 \le \omega^2 \le M^2_{\parallel,\bot}$. This property has been verified in the previous works\cite{toll,adl}. Therefore one can effectively set the integration range of Eq. (\ref{mellin}) as from $y=0$ to $y=\infty$. Now, it is easily seen that the derivatives of $\Pi_{\parallel,\bot}$ at the zero energy are proportional to the Mellin transform of $\kappa_{\parallel,\bot}
\cdot y^{-1/2}\equiv \kappa_{\parallel,\bot}\cdot \omega/M_{\parallel,\bot}$.
Once the l.h.s. of Eq. (\ref{mellin}) 
is calculated, the absorption coefficients $\kappa_{\parallel,\bot}$ can
be determined by the inverse Mellin transform. 

To calculate 
$\Pi_{\parallel,\bot}$ and their derivatives, we begin with the proper-time
representation of vacuum polarization function $\Pi_{\mu\nu}$ in a 
background magnetic field\cite{TS}:
\begin{eqnarray}
\Pi_{\mu\nu}(q)&=&-{e^3B\over (4\pi)^2}\int_0^{\infty}ds
\int_{-1}^{+1} dv \{e^{-is\phi_0}[(q^2g_{\mu\nu}-
q_{\mu}q_{\nu})N_0 \nonumber \\
&-&(q_{\parallel}^2g_{\parallel\mu\nu}-
q_{\parallel\mu}q_{\parallel\nu})N_{\parallel} 
+(q_{\bot}^2g_{\bot\mu\nu}-
q_{\bot\mu}q_{\bot\nu})N_{\bot}]\nonumber \\
&-&e^{-ism_e^2}(1-v^2)(q^2g_{\mu\nu}-
q_{\mu}q_{\nu})\},
\label{proper_t}
\end{eqnarray}
where
\begin{equation}
\phi_0=m_e^2-{1-v^2\over 4}q_{\parallel}^2-{\cos(zv)-\cos(z)\over 2z\sin(z)}
q_{\bot}^2
\end{equation}
with $z=eBs$, and
\begin{eqnarray}
N_0&=&{\cos(zv)-v\cot(z)\sin(zv)\over \sin(z)},\nonumber \\
N_{\parallel}&=&-\cot(z)\left(1-v^2+{v\sin(zv)\over \sin(z)}\right)
+{\cos(zv) \over \sin(z)},\nonumber \\
N_{\bot}&=&-{\cos(zv)\over \sin(z)}
+{v\cot(z)\sin(zv)\over \sin(z)}+2{\cos(zv)-\cos(z)\over \sin^3(z)},
\end{eqnarray}
To construct $\Pi_{\parallel,\bot}$ from $\Pi_{\mu\nu}$, we note that
only the structures proportional to $N_{\parallel}$ and $N_{\bot}$ 
contribute to $\Pi_{\parallel,\bot}$. Since we only concern
with the limit $\omega\ll m_e$ and $B< B_c$, $\Pi_{\parallel,\bot}$ can be
expanded in a series\footnote{In fact, we do not need 
Eq. (\ref{proper_t}) to obtain such an expansion. A convenient 
weak-field expansion technique applicable to the current problem 
has been developed in Ref.\cite{FIELD}.} of $\omega$ and $B$:
\begin{equation}
\Pi_{\parallel,\bot}=\sum_{n=1}^{\infty}{2\alpha m_e^2\over \pi}
\left({\omega^2\sin^2\theta B^2\over 3m_e^2 B_c^2}\right)^n
{\Gamma(3n-1)\Gamma^2(2n)\over \Gamma(n)
\Gamma(4n)}\left({6n+1, 3n+1\over 4n+1}\right)+\cdots ,
\end{equation}
where the neglected terms are of the order $(\omega^2B^2
\sin^2\theta/m_e^2 B_c^2)^n
(B/B_c)^2$.
Taking the derivatives of $\Pi_{\parallel,\bot}$, we arrive at
\begin{equation}
{1\over n!}\left({d^n\over d(\omega^2)^n}
\Pi_{\parallel,\bot}\right)\Big{\vert}_{\omega^2=0}
={2\alpha m_e^2\over \pi}\left({B^2\sin^2\theta\over 3B_c^2m_e^2}\right)^n
{\Gamma(3n-1)\Gamma^2(2n)\over \Gamma(n)
\Gamma(4n)}\left({6n+1, 3n+1\over 4n+1}\right)+\cdots
\label{diff}
\end{equation} 

Combining the above equation and Eq. (\ref{mellin}), the absorption
coefficients $\kappa_{\parallel,\bot}$ can be written in terms  
of inverse Mellin transform:
\begin{eqnarray}
\kappa_{\parallel}&=&{\alpha m_e^2\over i\pi\omega}
\int_{-i\infty+a}^{+i\infty+a} ds {(\lambda^{'})}^{2s}
{\Gamma(3s)\Gamma^2(2s)\over \Gamma(s)\Gamma(4s)}{1\over 3s-1}
\times {6s+1\over 4s+1},\nonumber \\
\kappa_{\bot}&=&{2\alpha m_e^2\over i\pi\omega}{1\over 1+\sqrt{1+2B/B_c}}
\int_{-i\infty+a}^{+i\infty+a} ds {(\lambda^{''})}^{2s}
{\Gamma(3s)\Gamma^2(2s)\over \Gamma(s)\Gamma(4s)}{1\over 3s-1}
\times {3s+1\over 4s+1},\nonumber \\
\label{abso}
\end{eqnarray}  
where $a$ is any real number greater than $1/3$; while $\lambda^{'}=
(\omega\sin\theta B/ \sqrt{3}m_e B_c)$ and
$\lambda^{''}=\lambda^{'}\cdot (1+\sqrt{1+2B/B_c})/2$.
At this moment, we only concern with the leading magnetic-field 
effect to the absorption coefficients, hence
we may set $(1+\sqrt{1+2B/B_c})\to 2$ and
$\lambda^{''}\to \lambda^{'}$ in $\kappa_{\bot}$. 
Numerically we find no distinctions between our results and the results of
Tsai and Erber\cite{TE}. 
For $\lambda'=10$, we have
$\kappa_{\parallel}=7.2\times (\alpha m_e^2/ \pi\omega)$ while 
$\kappa_{\bot}=4.4\times (\alpha m_e^2/ \pi\omega)$. For $\lambda'=100$, 
the above absorption coefficients become $42\times (\alpha m_e^2/ \pi\omega)$
and $28\times (\alpha m_e^2/ \pi\omega)$ respectively.  
In the current approximation, $\kappa_{\parallel}$
is always greater than $\kappa_{\bot}$. For a high-energy photon, i.e.,
$\lambda'\gg 1$, we have $\kappa_{\parallel}/\kappa_{\bot}=1.5$. 
This is already reflected in the above case with $\lambda'=100$. For the
low-energy photon, $\lambda'\ll 1$, we find 
$\kappa_{\parallel}/\kappa_{\bot}=2$.
The numerical agreement between Eq. (\ref{abso}) and the result of 
Ref.\cite{TE}, as shown in Eq. (\ref{bessel}), is 
not a coincidence. We shall verify shortly that both expressions are
equivalent by comparing their infinite sequences of moments. 

As mentioned 
earlier, in order to establish the 
equivalence of their result with those of the previous works, 
the authors of Ref.\cite{TE} computed the {\it moments} 
$C_n^{(1)}=\int_0^{\infty}d\chi \chi^n T(\chi)$ with $T(\chi)=1/2\cdot
(T_{\parallel}(\chi)+T_{\bot}(\chi))$ and $\chi=(4m_eB_c/3\omega B)$.
The superscript $(1)$ is used to denote the set of {\it moments} computed
from the functions $T_{\parallel,\bot}$ given in Ref. \cite{TE}. The 
superscript $(0)$ will then be used for denoting the {\it moments}
computed from our results for $T_{\parallel,\bot}$.
Without taking the average, we obtain the {\it moments}
of each individual function $T_{\parallel}$ and $T_{\bot}$, which we 
denote as $C_{n}^{(1)}(\parallel)$ and $C_{n}^{(1)}(\bot)$ respectively.
We obtain
\begin{equation}
C_{n}^{(1)}(\parallel,\bot)=2^{-2}(3/\pi)^{1/2}{(6n+14,3n+8)\over 2n+5}
{\Gamma(n/2+2/3)\Gamma(n/2+4/3)\Gamma(n+2)\over \Gamma(n+5/2)},
\label{momte}
\end{equation}
To compare  $C_{n}^{(1)}(\parallel,\bot)$ with the  
{\it moments} pertinent to Eq. (\ref{mellin}), i.e., 
$D^{(0)}_n \equiv \int_{0}^{\infty}dy\cdot y^{n-1}\cdot \left(T_{\parallel
,\bot}(y) y^{-1/2}\right)$, we employ the relation        
\begin{equation}
D_n^{(0)} (\parallel, \bot)= 2 \left 
( {3 B \over 2 B_c \sin \theta} \right )^
{2n-1}C^{(0)}_{2n-2}(\parallel, \bot),
\label{d0c0}
\end{equation}  
where $C^{(0)}_{2n-2}(\parallel, \bot)$ is defined in the same way as
$C^{(1)}_{2n-2}(\parallel, \bot)$. Our first goal is to show that
$C^{(0)}_{2n-2}(\parallel, \bot)=C^{(1)}_{2n-2}(\parallel, \bot)$.
In fact, this identity can be established by combining the relation
\begin{equation}
D_n^{(0)} (\parallel, \bot)= {2^{2n+1} \over 3^n}
\left ( {B \over B_c} \right) ^{2n-1} 
{\Gamma (3n-1) \Gamma(2n) \Gamma(2n) \over  
\Gamma(n)\Gamma(4n) }\cdot { (6n+1, 3n+1) \over 4n+1},
\end{equation} 
derived from Eqs. (\ref{mellin}) and (\ref{diff}), with Eqs. 
(\ref{momte}), (\ref{d0c0}) and
the identity
\begin{equation}
{ \Gamma ( n-1/3) \Gamma( n+1/3) \over  
\Gamma(2n+1/2) } =  {  2^{4n} \sqrt{\pi} \over  3^{3n-3/2} } 
\times { \Gamma (3n-1) \Gamma ( 2n) 
\over \Gamma (n) \Gamma (4n ) }.
\end{equation}   
Now that we have shown 
$C^{(0)}_{2n-2}(\parallel, \bot)=C^{(1)}_{2n-2}(\parallel, \bot)$,
we obtain the identity
$D_n^{(0)} (\parallel, \bot)=D_n^{(1)} (\parallel, \bot)$
where $D_n^{(1)} (\parallel, \bot)$ is 
given by the r.h.s. of Eq. (\ref{d0c0})
with $C^{(0)}_{2n-2}(\parallel, \bot)$ replaced by  
$C^{(1)}_{2n-2}(\parallel, \bot)$. Since both $D_n$'s are identical, 
one can show that 
the absorption coefficient
derived from our approach, Eq. (\ref{abso}), is equivalent to 
the result of Ref.\cite{TE} given by Eqs. (\ref{bessel}) and (1), provided that
\begin{equation}
\sum_n  D_n^{-1/2n} \to \infty
\end{equation}
according to the Carleman's theorem \cite{carleman}. 
Indeed, this is true since 
\begin{equation}
\sum_n  D_n^{-1/2n} \to \sum_n  {1 \over  n} \to \infty \ {\rm as} \
D_n \sim  {9 \over 8} ({ 2\pi \over 3} )^{1/2}
n^{2n-3/2} e^{2-2n}.
\end{equation}  
Therefore, we have proven that the leading-order results of our 
approach agree 
with the results of Ref.\cite{TE}.
In addition, as one can see from the r.h.s. of Eq. (\ref{d0c0}), only 
the even moments defined by Tsai and Erber
are relevant to the physics of pair-production in a 
background magnetic field.

We like to point out the differences between our approach and
the approach of Ref.\cite{TE}. Tsai and Erber begin with $\Pi_{\mu\nu}$
given in Eq. (\ref{proper_t}) and evaluate the imaginary part of
$\Pi_{\mu\nu}$ for $\omega\equiv q^0$ greater than the pair production
threshold. They arrive at the asymptotic result, Eq. (\ref{bessel}), 
in the limit 
$B\ll B_c$ and $\omega\sin\theta \gg 2m_e$. However, their approach 
does not provide an estimate of possible corrections as $B$ and $\omega$ 
deviate from the above limit.    
Our approach has an advantage in that it treats the magnetic-field 
effects perturbatively for $B<B_c$. 
In Eq. (\ref{diff}), the $n$-th derivative
of the vacuum-polarization function $\Pi_{\parallel,\bot}$ 
is expanded in powers of $B^2/B_c^2$.  
Hence the absorption coefficient, which is related to the derivatives of
$\Pi_{\parallel,\bot}$ by an inverse Mellin transform, can also be 
written in powers of $B^2/B_c^2$. In this way, we are able to compute
the absorption coefficient even for $B$ comparable to $B_c$. As for 
the low energy regime near the pair production
threshold, $\omega \sin\theta \agt 2m_e$, the quantum effects due to
the magnetic field become important. Namely, for given $\omega$ and $B$,
the momenta of $e^+$ and $e^-$ along the magnetic-field direction
can only take discrete values, and consequently the absorption coefficients 
$\kappa_{\parallel,\bot}$ contain resonant peaks. The spacing 
of these peaks increases as $\omega \sin\theta$ gradually decreases to the 
pair-production threshold $2m_e$. A detailed study of this threshold
behavior has been initiated by Daugherty and Harding
\cite{DH}. In Fig. 6 of Ref.\cite{DH}, it is shown
that the threshold behavior is non-negligible for $\xi\equiv \omega^2 B_c
/2m_e^2 B< 10^3$ with $\sin^2\theta=1$. For a general $\theta$, the
relevant parameter becomes  $\xi^{'}=\omega^2\sin^2\theta B_c/2m_e^2 B$.  
It should be understood that our result as well as the result of 
Ref.\cite{TE} are applicable for a large $\xi(\xi^{'})$ where
the threshold effect is not significant.
          
We have mentioned that our results for $\kappa_{\parallel,\bot}$ are 
written as power series in $(B/B_c)^2$. It is important to compute
the next-to-leading corrections.
Let us begin by computing the next-to-leading magnetic-field corrections
to $\Pi_{\parallel}$ and its derivatives. To do this we perform a weak-field
expansion with respect to the exponent $\phi_0$ and the amplitude 
$N_{\parallel}$, along with a rotation of the integration 
contour $s\to -is$:
\begin{eqnarray}
\phi_0&=&m_e^2-{\omega^2\sin^2\theta\over 48}z^2 (1-v^2)^2 (1-{1\over 30}
(3-v^2)z^2+\cdots),\nonumber \\
-iN_{\parallel}&\to &\left[{\cosh(zv)\over \sinh(z)}
-{\cosh(z)\over \sinh(z)}(1-v^2+{v\sinh(zv)\over \sinh(z)})\right]
\nonumber \\
&=&-{z\over 6}(1-v^2)(3-v^2)+{z^3\over 360}(1-v^2)(15-2v^2+3v^4)+\cdots.
\end{eqnarray}
Hence the next-to-leading (NL) corrections to the derivatives of 
$\Pi_{\parallel}$
reads:
\begin{eqnarray}
{1\over n!}\left({d^n\over d(\omega^2)^n}
\Pi^{NL}_{\parallel}\right)\Big{\vert}_
{\omega^2=0}&=&{-2\alpha m_e^2\over 5\pi}
({B\over B_c})^2\left({B^2\sin^2\theta\over 3B_c^2m_e^2}\right)^n
{n\Gamma(3n)\Gamma^2(2n)\over \Gamma(n)
\Gamma(4n)}\nonumber \\
&\times &{(3+2n+24n^2+36n^3)\over (4n+1)(4n+3)}.
\end{eqnarray}
Using Eq. (\ref{mellin}), and applying the inverse Mellin transform, we
arrive at
\begin{equation}
\kappa_{\parallel}^{NL}={-\alpha m_e^2\over 10 i\pi\omega}({B\over B_c})^2
\int_{-i\infty+b}^{+i\infty+b} ds {(\lambda^{'})}^{2s}
{\Gamma(3s)\Gamma^2(2s)s\over \Gamma(s)\Gamma(4s)}
\times {(3+2s+24s^2+36s^3)\over (4s+1)(4s+3)},
\end{equation}
where $b$ can be chosen to be any positive number.
Numerically, for $(B/B_c)^2=0.1$ and $\lambda'=10$, we have
$\kappa^{NL}_{\parallel}=-1.5\times 10^{-3}\times
(\alpha m_e^2/ \pi\omega)$. We note that $\xi^{'}\approx 10^4$
for the current values of $B$ and $\lambda^{'}$.
In this case 
$\vert \kappa^{NL}_{\parallel}/\kappa_{\parallel}\vert< 0.1\%$. 
For the same magnetic-field strength with $\xi^{'}=10^3$
($\lambda'\approx 3$),
we obtained $\kappa_{\parallel}=2.0\times
(\alpha m_e^2/ \pi\omega)$ and 
$\kappa^{NL}_{\parallel}=7.7\times 10^{-4}\times
(\alpha m_e^2/ \pi\omega)$.
The ratio $r\equiv \vert
\kappa^{NL}_{\parallel}/\kappa_{\parallel}\vert$ remains to be less than 
$0.1$\% in this case. Hence in the energy regime that the 
quantum effects of the magnetic field is not essential, the 
subleading contribution to the absorption coefficient, 
$\kappa^{NL}_{\parallel}$, is rather suppressed.
If we extrapolate our analysis down to the energy of 
pair-production threshold 
$\omega\sin\theta=2m_e$
while maintaining $(B/B_c)^2=0.1$, i.e., $\lambda'=0.35$, we find 
$\kappa_{\parallel}=6.5\times 10^{-3}\times
(\alpha m_e^2/ \pi\omega)$ and 
$\kappa^{NL}_{\parallel}=-2.2\times 10^{-3}\times
(\alpha m_e^2/ \pi\omega)$. It is interesting to see that
$\kappa^{NL}_{\parallel}$ is of the same order of magnitude as 
the leading contribution. This reflects the limitation of 
our approach and that of Tsai and Erber near the 
pair-production threshold. 

The next-to-leading correction to $\kappa_{\bot}$ is calculated in a 
similar way. 
The expansion of $\phi_0$ proceeds as before
while 
\begin{eqnarray}
-iN_{\bot}&\to &\left[-{\cosh(zv)\over \sinh(z)}
+{v\cosh(z)\sinh(zv)\over \sinh^2(z)}-{2(\cosh(zv)-\cosh(z))\over 
\sinh^3(z)}\right]\nonumber \\
&=&-{z\over 12}(1-v^2)(3+v^2)+{z^3\over 180}(1-v^2)(15-6v^2-v^4)+\cdots.
\end{eqnarray}
Then the next-to-leading corrections to the derivatives of $\Pi_{\bot}$ are
\begin{eqnarray}
{1\over n!}\left({d^n\over d(\omega^2)^n}
\Pi^{NL}_{\bot}\right)\Big{\vert}_{\omega^2=0}&=&{-2\alpha m_e^2\over 5\pi}
({B\over B_c})^2\left({B^2\sin^2\theta\over 3B_c^2m_e^2}\right)^n
{n\Gamma(3n)\Gamma^2(2n)\over \Gamma(n)
\Gamma(4n)}\nonumber \\
&\times &{(3+39n+60n^2+18n^3)\over (4n+1)(4n+3)}.
\end{eqnarray}
Applying the inverse Mellin transform, we obtain
\begin{equation}
\kappa_{\bot}^{NL}={-\alpha m_e^2\over 10 i\pi\omega}({B\over B_c})^2
\int_{-i\infty+c}^{+i\infty+c} ds {(\lambda^{'})}^{2s}
{\Gamma(3s)\Gamma^2(2s)s\over \Gamma(s)\Gamma(4s)}
\times {(3+39s+60s^2+18s^3)\over (4s+1)(4s+3)},
\end{equation}
where $c$ can be chosen as any positive number, and, 
to isolate the $O(B^2/B_c^2)$ corrections, we have made 
the identifications $\lambda^{''}\to\lambda^{'}$ and 
$(1+\sqrt{1+2B/B_c})\to 2$.
Numerically, for $(B/B_c)^2=0.1$ and $\lambda'=10$, we have
$\kappa_{\bot}=4.4\times (\alpha m_e^2/ \pi\omega)$ while
$\kappa^{NL}_{\parallel}=1.3\times 10^{-3}\times
(\alpha m_e^2/ \pi\omega)$.
Similar to the $\kappa_{\parallel}$ case,
$\vert \kappa^{NL}_{\bot}/\kappa_{\bot}\vert< 0.1\%$ for
$\lambda^{'}=10$. For $\lambda'=3$ which corresponds to
$\xi^{'}\approx 10^3$, we find $\kappa_{\bot}=1.2\times 
(\alpha m_e^2/ \pi\omega)$ and 
$\kappa^{NL}_{\bot}=-4\times 10^{-3}\times
(\alpha m_e^2/ \pi\omega)$. In this case 
$\vert \kappa^{NL}_{\bot}/\kappa_{\bot}\vert\approx 0.3\%$.
We observe again that the ratio, $\vert 
\kappa^{NL}_{\bot}/\kappa_{\bot}\vert$, grows rapidly to $60\%$ at the
energy of pair-production threshold 
with $\kappa^{NL}_{\bot}$ being negative. 

>From the above next-to-leading order calculations, 
it is quite evident that 
the $O(B^2/B_c^2)$-corrections to $\kappa_{\parallel}$
and $\kappa_{\bot}$ are both rather insignificant. However, one should be
reminded that there are still
$O(B/B_c)$- corrections to $\kappa_{\bot}$ as shown in Eq.(\ref{abso}). 
Without making the identifications $\lambda^{''}\to\lambda^{'}$ and 
$(1+\sqrt{1+2B/B_c})\to 2$, we have $\kappa_{\bot}=3.8
\times (\alpha m_e^2/ \pi\omega)$ for $\lambda^{''}=10\times  
(1+\sqrt{1+2B/B_c})/2$ (i.e., $\lambda'=10$) with 
$(B/B_c)^2=0.1$. 
We recall that $\kappa_{\bot}=4.4\times (\alpha m_e^2/ \pi\omega)$
if the above identifications are made. Hence the 
$O(B/B_c)$-correction reduces $\kappa_{\bot}$ by about $14\%$ at the
current $\omega$ and $B$. Essentially, the $O(B/B_c)$-corrections 
to $\kappa_{\bot}$ relative to $\kappa_{\parallel}$ are kinematic in
nature. They are due to the differences 
in pair-production threshold resulting
from different polarization states of the decaying photons.
For a photon with the $\parallel$-polarization, 
the pair-production threshold 
corresponds to both $e^+$ and $e^-$ being in the ground state. 
On the other hand, for a photon with the $\bot$-polarization, 
either $e^+$ or $e^-$ must be in the
first excited state at the pair-production threshold \cite{toll}.     

Since the $O(B^2/B_c^2)$-corrections to $\kappa_{\parallel,\bot}$ can be
neglected, Eq. (\ref{abso}) are accurate expressions for 
photon absorption coefficients even to a magnetic-field strength
comparable to $B_c$. In Fig. 2, we plot $\kappa_{\parallel,\bot}$ as
functions of $\omega$ with $\sin^2\theta=1$ and $(B/B_c)^2=0.1$.
Here our result for $\kappa_{\parallel}$ is identical to that of
Tsai and Erber, whereas our result for $\kappa_{\bot}$ contains the
kinematic corrections which were not taken into accounts in 
previous works.
As can be seen from Fig. 2, both $\kappa_{\parallel}$ and 
$\kappa_{\bot}$ are rather insensitive to $\omega$ once 
$\omega$ surpasses $10$ MeV. Hence, for a sufficiently large $\omega$,
the kinematic corrections to $\kappa_{\bot}$ mainly reside in the 
factor $1/(1+\sqrt{1+2B/B_c})$ rather than in $\lambda^{''}$.   It should be 
noted that the applicability of our result is determined by the
parameter $\xi^{'}=\omega^2\sin^2\theta B_c/2m_e^2 B$\cite{DH}. 
For the parameter  
set in Fig. 2, i.e., $(B/B_c)^2=0.1$ and $\sin^2\theta=1$, we have 
$\omega\approx 25 m_e
= 12.5$ MeV for $\xi^{'}=10^3$. 
Hence, one expects the plots in Fig. 2 to be reliable 
for $\omega> 10$ MeV.      

\vspace{1cm}
\begin{figure}
  \unitlength 1mm
   \begin{center}
      \begin{picture}(25,100)
     \put(-70,30) {\epsfig{file=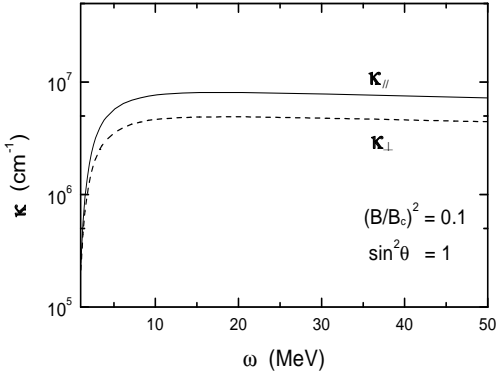,width=7cm,height=7cm}}
      \end{picture}
   \end{center}
\vspace{-3cm}
\caption{The photon absorption coefficients as functions of the photon energy
with $\sin^2\theta=1$ and $(B/B_c)^2=0.1$.}
\label{fig2}
\end{figure} 

At this point, one might conclude that our approach can not describe the threshold behavior of the absorption coefficient. This is in fact not true. 
We should stress that it is the weak-field expansion of Eq. (\ref{diff}) that spoils the threshold behavior of the absorption coefficient, despite such an expansion is useful for computing the absorption coefficient at much higher energies.    
To see this, it is instructive to review how Eq. (\ref{diff})
is derived. In this note, we obtained Eq. (\ref{diff}) 
by expanding Eq. (\ref{proper_t}) directly. However,
one could also derive Eq. (\ref{diff}) by expanding the internal fermion
propagators in $\Pi_{\parallel,\bot}$ in powers of $eB$. 
This expansion has been derived in Eq. (47) of Ref. \cite{FIELD}. 
In that equation, one can see that the weak-field expansion mixes contributions coming from different Landau levels. 
Hence the threshold behaviors of $\Pi_{\parallel,\bot}$ at any given Landau level are spoiled by the weak-field expansion. 
In fact, the absorption coefficients $\kappa_{\parallel,\bot}$ calculated from $\Pi_{\parallel,\bot}$ do not vanish below the pair-production threshold, i.e., $0 \le \omega^2 \le M^2_{\parallel, \bot}$. 
This can be seen by performing the integration in Eq. (\ref{abso}). 
Such a behavior is again an artifact caused by the weak field expansion performed near the pair-production threshold. Indeed, as mentioned before, $\kappa_{\parallel,\bot}^{NL}$ are comparable to $\kappa_{\parallel,\bot}$ with an opposite sign in this energy range. The cancellation of these two contributions is consistent with the fact that the absorption coefficients should vanish below the pair-production threshold \cite{toll,adl}.
  
To recover the threshold behavior, one may compute $\Pi_{\mu\nu}$
using electron propagators in the Furry picture\cite{FY,KS}:
\begin{equation}
 S^B_F(x',x)_{\alpha\beta}=\sum_{n=0}^{\infty}\int 
{d\omega dq^2dq^3\over (2\pi)^3}
{\exp(-i\omega(t'-t)+iq^2(x^{'2}-x^2)+iq^3(x^{'3}-x^3))
\over q_{\parallel}^2-m_e^2-2neB+i
\epsilon}(S_{n;\omega,q^2, q^3})_{\alpha\beta},
\end{equation}
where $q_{\parallel}^2=(q^0)^2-(q^3)^2$, $n$ is the quantum number for 
the Landau level, and
$S_{n;\omega,q^2, q^3}$ is a $4\times 4$ matrix in the spinor space. 
In this form, all the 
poles of the propagator appear explicitly, and the 
threshold behavior of $\Pi_{\mu\nu}$
is preserved throughout the calculation\cite{KLT}.
Clearly, a calculation using the Furry-picture 
propagators compliments the weak-field expansion technique 
we have 
been discussing so far. The former produces a correct threshold-behavior 
of the absorption coefficient but becomes unpractical at larger energies,
since, in such a case, contributions from a 
great number of Landau levels has to be summed over.
Nevertheless, in the scenario that $B\gg B_c$, 
it is convenient to use the 
Furry-picture propagators because the available Landau levels for pair
production to occur are significantly reduced. 

In conclusion, we have developed a new method for computing the 
photon absorption coefficient in a strong background 
magnetic field $B\alt B_c$. 
Disregarding the next-to-leading magnetic-field corrections, 
our $\kappa_{\parallel}$
is identical to that obtained by Tsai and Erber\cite{TE}. 
Although Tsai and Erber 
derived $\kappa_{\parallel}$ under the assumption 
$B\ll B_c\equiv m_e^2/e$ and 
$\omega \sin\theta \gg 2m_e$, we have been able to show that 
such a result is in fact
accurate for $B$ comparable to $B_c$, provided $\omega$ ($\xi^{'}$) is
large enough. For $\kappa_{\bot}$, our result
differs from that obtained in Ref.\cite{TE}.  
In this regard, we have identified certain $O(
B/B_c)$ corrections to $\kappa_{\bot}$ which are 
kinematic in nature. We also pointed out 
that our approach may be extended to lower photon energies 
near the pair-production threshold, 
so long as we calculate $\Pi_{\mu\nu}$ with 
electron propagators in the Furry picture\cite{KS}. 
For a supercritical magnetic field $B\gg B_c$, we argued that it is 
convenient to use the Furry-picture propagators. 
As a closing, we like to emphasize that a better
understanding of the current process  
is crucial for determining the photon attenuation properties in
highly magnetized pulsars\cite{BH98}.   

\acknowledgments 
We thank H.-K. Chang for
bringing Ref.\cite{DH}
to our attentions. This work is supported in part 
by the National Science Council under
grant numbers NSC89-2112-M009-001 and NSC89-2112-M009-035.


\begin{thebibliography}{99}
%
\bibitem{STU}
P. A. Sturrock, ApJ {\bf 164}, 529, 1971.
%
\bibitem{BH}
M. G. Baring and A. K. Harding, ApJ {\bf 482}, 372, 1997.
%
\bibitem{toll}
J. S. Toll, Ph.D. thesis, Princeton Univ., 1952 (unpublished).
%
\bibitem{adl}
S. L. Adler, Ann. Phys. (N.Y.) 67, 599, 1971.
%
\bibitem{TE}
W.-y. Tsai and T. Erber, Phys. Rev. D {\bf 10}, 492, 1974. 
%
\bibitem{SCH}
J. Schwinger, Phys. Rev. {\bf 82}, 664, 1951.
%
\bibitem{SVZ}
M. A. Shifman, A. I. Vainstein, and V. I. Zakharov, Nucl. Phys. B {\bf 147},
385, 1979.
%
\bibitem{TS}
W.-y. Tsai, Phys. Rev. D {\bf 10}, 2699, 1974.
%
\bibitem{FIELD}
T.-K. Chyi, C.-W. Hwang, W.-F. Kao, G.-L. Lin, K.-W. Ng and J.-J. Tseng,
hep-th/9912134, to appear in Phys. Rev. D.
\bibitem{carleman}
T. Carleman, {\it Les fonctions quasi-analytiques } (Herman, Paris, 1926).
%
\bibitem{DH}
J. K. Daugherty and A. K. Harding, ApJ {\bf 273}, 761, 1983.
%
\bibitem{FY}
W. H. Furry, Phys. Rev. {\bf 81}, 115, 1951.
\bibitem{KS}
M. Kobayashi and M. Sakamoto, Prog. Theor. Phys., {\bf 70}, 1375, 1983.
%
\bibitem{KLT}
W.-F. Kao, G.-L. Lin and J.-J. Tseng, work in progress.
\bibitem{BH98}
M. G. Baring and A. K. Harding, ApJ {\bf 507}, L55, 1998.
\end{thebibliography}
\end{document}